\documentclass{aa}  

\usepackage{graphicx}
\usepackage{txfonts}
\usepackage[usenames]{color}
\usepackage{caption}
\usepackage{subcaption}
\usepackage{hyperref}
\usepackage{natbib}
\usepackage[normalem]{ulem}
\usepackage[dvipsnames]{xcolor}

\hypersetup{colorlinks=true,linkcolor=blue,citecolor=blue,urlcolor=blue}

\definecolor{green}{rgb}{0.,1.,0.}
\definecolor{greenMM}{rgb}{0.,0.8,0.7}
\definecolor{orange}{rgb}{1,0.5,0}
\colorlet{LightRubineRed}{RubineRed!70!}

\def\aap{A\&A}

\def\apjl{ApJL}

\begin{document} 


\title{Imaging the expanding knotty structure in the close environment of the LBV star $\eta$ Carinae}

   \author{F.Millour
          \inst{1}
          \fnmsep\thanks{Based on observations made with European Southern Observatory
(ESO) telescopes at Paranal Observatory in Chile, under program ID
096.D-0252(A).}
          \and
          E. Lagadec\inst{1}
          \and 
          M. Montarg\`es\inst{2}
          \and
          P. Kervella\inst{3}
          \and
          A. Soulain\inst{1
          \fnmsep\thanks{Now at Sydney Institute for Astronomy (SIfA), School of Physics, The University of Sydney, NSW 2006, Australia}}
          \and\\
          F. Vakili\inst{1}
          \and
          R. Petrov\inst{1}
          \and
          G. Weigelt\inst{4}
          \and
          J. Groh\inst{5}
          \and
          N. Smith\inst{6}
          \and
          A. Mehner\inst{7}
          \and
          H. M. Schmid\inst{8}
          \and
          J. Ramos\inst{9}
          \and
          O. Moeller-Nillson\inst{9}
          \and
          R. Roelfsema\inst{10}
          \and
          F. Rigal\inst{10}
          }

   \institute{Universit\'e C\^ote d'Azur, Observatoire de la C\^ote d'Azur, CNRS, Lagrange, France\\
              \email{fmillour@oca.eu}
         \and
         Institute of Astronomy, KU Leuven, Celestijnenlaan 200D B2401, 3001 Leuven, Belgium
          \and
          LESIA (UMR 8109), Observatoire de Paris, PSL Research University, CNRS, UPMC, Univ. Paris-Diderot, 5 place Jules Janssen, 92195, Meudon, France
          \and 
             Max Planck Institute for Radio Astronomy, Auf dem Hügel 69, 53121, Bonn, Germany
          \and
            Trinity College Dublin, The University of Dublin, Dublin 2, Ireland
          \and
             Steward Observatory, University of Arizona, 933 N. Cherry Ave., Tucson, AZ 85721, USA
              \and
ESO - European Southern Observatory, Alonso de C\'{o}rdova 3107, Vitacura, Casilla 19001, Santiago de Chile, Chile
          \and
              Institute for Particle Physics and Astrophysics, ETH Zurich, Wolfgang-Pauli-Strasse 27, 8093 Zurich, Switzerland
          \and
             Max Planck Institute for Astronomy, K\"onigstuhl 17, D-69117 Heidelberg, Germany
          \and
              NOVA Optical Infrared Instrumentation Group, Oude Hoogeveensedijk 4, 7991 PD Dwingeloo, The Netherlands
             }


 
  \abstract
   {$\eta$~Car is one of the most massive stars in the Galaxy. It underwent a massive eruption in the 19th century, which produced the impressive bipolar Homunculus nebula now surrounding it. The central star is an eccentric binary with a period of 5.54\,years. Although the companion has not been detected directly, it causes time-variable ionization and colliding-wind X-ray emission.}
   {By characterizing the complex structure and kinematics of the ejecta close to the star, we aim to constrain past and present mass loss of $\eta$~Car.}
   {$\eta$~Car is observed with the extreme adaptive optics instrument SPHERE at the Very Large Telescope, using its polarimetric mode in the optical with the ZIMPOL camera. A spatial resolution of 20\,mas was achieved, i.e. very close to the presumed 13 mas apastron separation of the companion star.}
   {We detect new structures within the inner arcsecond to the star (2\,300\,au at a 2.3\,kpc distance). We can relate these structures to the eruption near 1890 by tracking their proper motions derived from our new images and  historical images over a 30\,years time span. Besides, we find a fan-shaped structure in the inner 200~au to the star in the H$\alpha$ line, that could potentially be associated with the wind collision zone of the two stars.}
   {}


\titlerunning{$\eta$~Car with SPHERE/VLT}
   \maketitle
%
%









\section{Introduction}\label{sec:intro}

$\eta$~Car\footnote{also known as HD~93308} is a luminous blue variable star located in the Carina constellation in the Southern hemisphere. It is a massive star at the end of its life, at the verge of exploding as a supernova. As such, it is a live laboratory of a pre-supernova system.

With a mass of about $100~M_\odot$ and a luminosity $L\sim5\times10^6~L_\odot$, the primary star of the $\eta$~Car binary is also one of the most massive stars in the Milky Way. The star is close to the Eddington limit, and far above the Humphreys-Davidson limit, 
and presumably, for this reason, it is unstable \citep{1979ApJ...232..409H}. It has a long and spectacular history of variability and outbursts \citep{1984SSRv...39..317V, 1997ARAandA..35....1D,  2011MNRAS.415.2009S}. $\eta$~Car is located at a distance of 2.3$\pm$0.1~kpc in the Trumpler 16 OB cluster in Carina \citep{1973ApJ...179..517W, 2006ApJ...644.1151S}. 

The star is surrounded by a bipolar nebula, known as the ``Homunculus nebula'', ejected during the 19$^{\rm th}$-century Great Eruption \citep[around 1841][]{1950ApJ...111..408G, 2001ApJ...548L.207M, 2017MNRAS.471.4465S}. The lobes of this nebula extend about 20$^{\prime\prime}$ on the sky ($4.5 \times10^4$~au), and show a complex and highly clumped density structure \citep{ 1998AJ....116.2443M}. The Homunculus is mainly a reflection nebula in the optical, and most of the mass resides in the bipolar shell \citep{2006ApJ...644.1151S}. Dust absorbs optical and UV flux from the central star, and then re-emits at infrared wavelengths, making $\eta$~Car one of the brightest sources in the sky from 10 to 20~$\mu$m \citep{1969ApJ...156L..45W,  1986ApJ...311..380H, 2003AJ....125.1458S, 2017ApJ...842...79M}.

\begin{figure*}[htbp]
\centering
  \hspace*{-1.cm}
\includegraphics[height=19cm, angle=0]{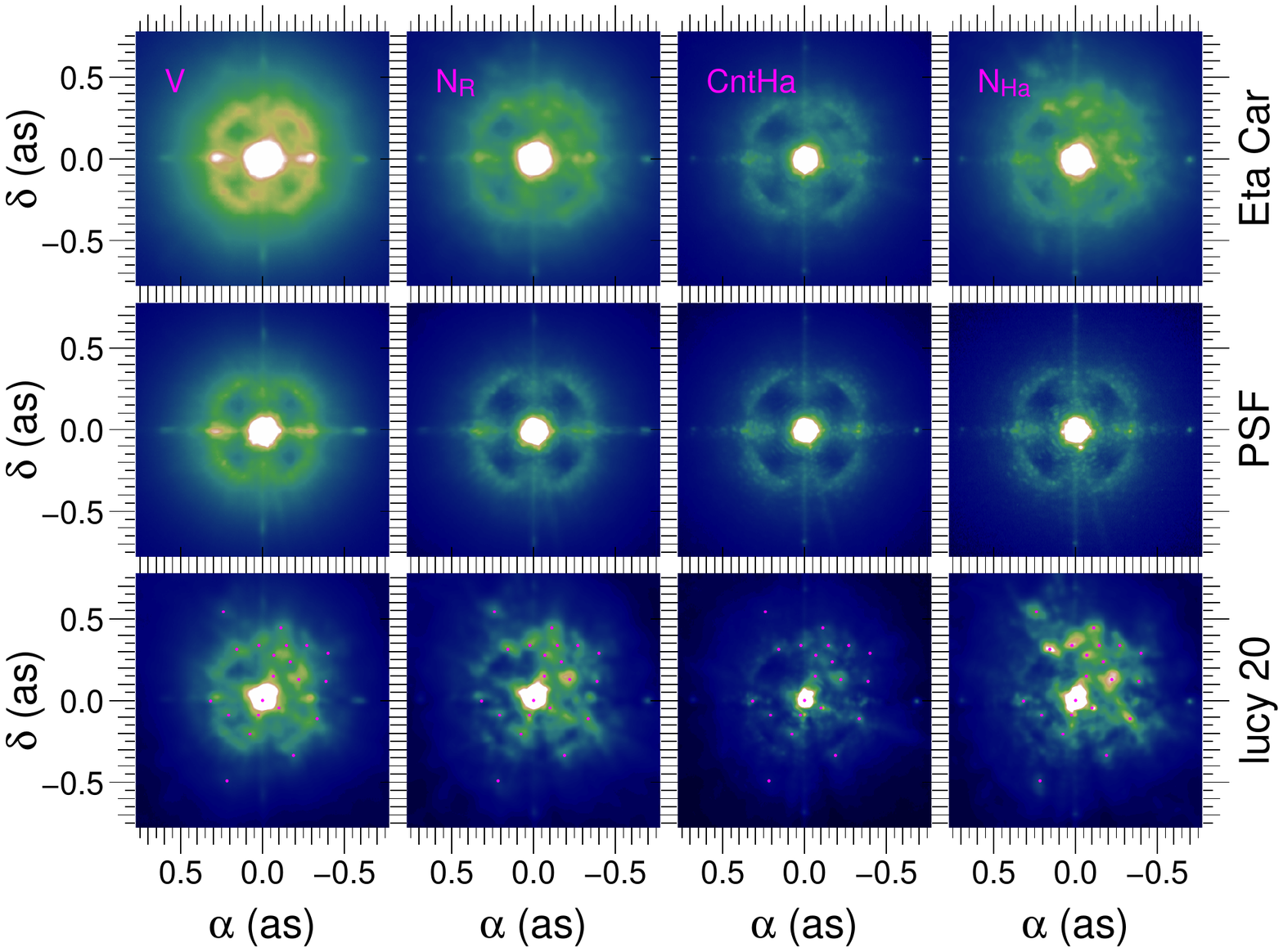}
  \vspace*{-4.5cm}
\caption{{\bf From left to right:} SPHERE images of $\eta$~Car in the V band ($\lambda=554$\,nm), R band ($\lambda=645.9$\,nm), H$\alpha$ continuum filter ($\lambda=644.9$\,nm), H$\alpha$ filter ($\lambda=656.34$\,nm). {\bf From top to bottom:} images of $\eta$~Car, of the PSF star HD89682, and finally Lucy-deconvolved images of $\eta$~Car with 20 iterations. We mark the identified features with pink dots on the Lucy 20 line.
}
  \label{fig:etacar_all}
\end{figure*}

Another smaller bipolar nebula, known as the "Little Homunculus", is nested inside the main Homunculus.  Based on its kinematics and size, the Little Homunculus is usually associated with a second minor eruption in the 1890s, nicknamed "the Lesser Eruption". With an ejected mass of only about 0.1 $M_{\odot}$ and a relatively slow expansion speed, the Lesser Eruption was far weaker than the Great Eruption, although it shared the same bipolar shape for its mass-loss geometry, oriented along the same polar axis as the larger Homunculus \citep{1999PASP..111.1124H, 2003AJ....125.3222I, 2005MNRAS.357.1330S}.

\begin{table}[bp]
\caption{\label{tab:SPHERE}Journal of observations with SPHERE ZIMPOL on February 19$^{\rm th}$, 2016. }
\vspace{-0.3cm}
\begin{center}
\begin{tabular}{lcccc}
\hline
\hline
Star & Filter &$t_{\rm exp}$ & Seeing & ND \\
 &  & (s) & (") & filter\\
\hline
$\eta$~Car & NHa / CnTHa&576  & 1.12 & ND1\\
$\eta$~Car & V / NR&960    & 1.20 & ND1\\
HD~89682 & NHa / CntHa&192& 1.40 & -\\
HD~89682 & V / NR&172.8& 1.46 & -\\
\hline
\hline
\end{tabular}
\end{center}
\end{table}

\begin{figure*}[htbp]
\centering
\vspace*{-2cm}
  \includegraphics[width=.95\linewidth, angle=0]{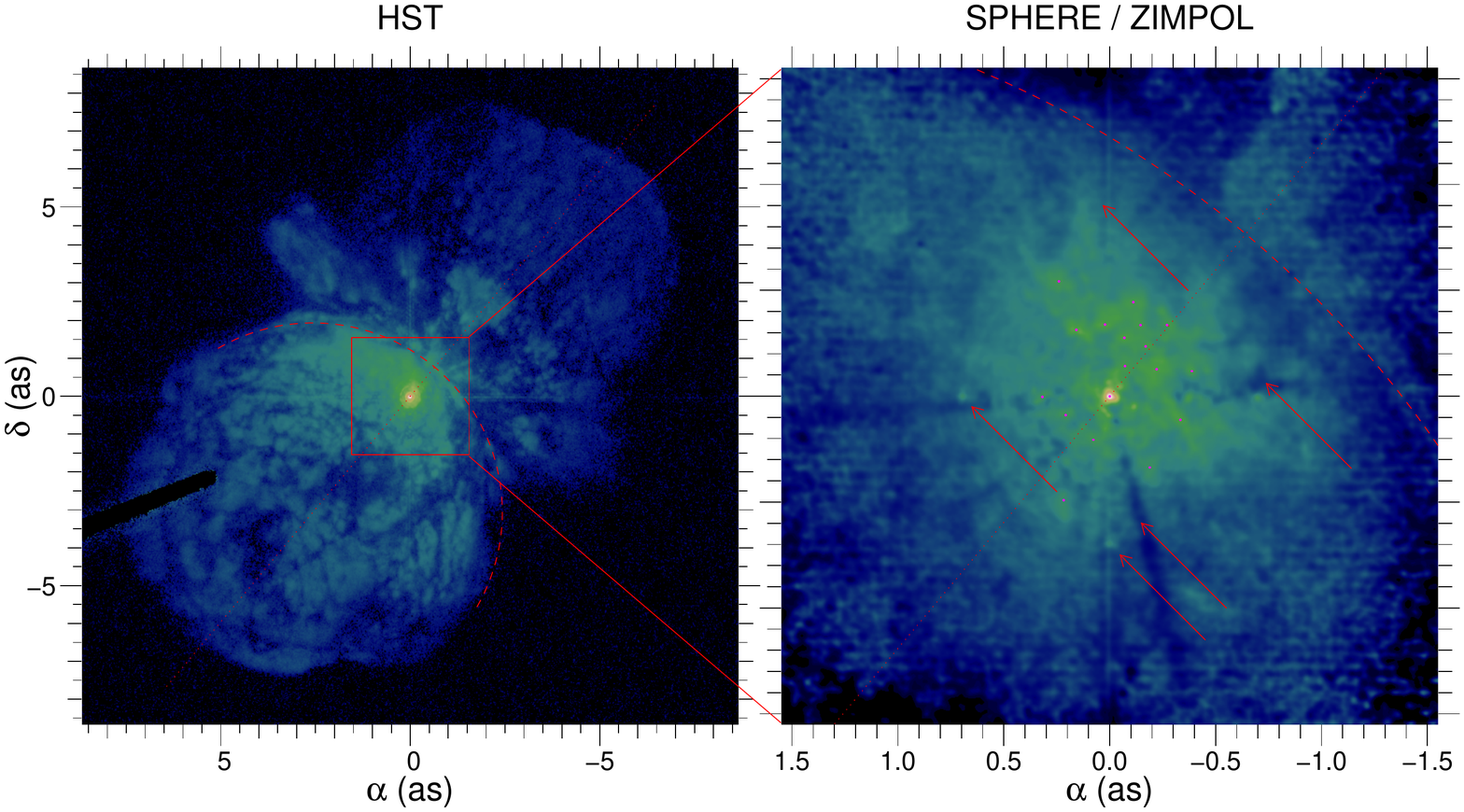}
  \vspace*{-1.cm}
  \caption{{\bf Left:} {\it HST/ACS} image of $\eta$~Car in H$\alpha$ taken in 2003. {\bf Right:} SPHERE ZIMPOL deconvolved image of $\eta$~Car in the H$\alpha$ line with the newly identified knot positions indicated (see text for details). We show an arc-like structure in dashed line, and deconvolution artefacts with arrows. We display here the hyperbolic arc-sinus of the intensity in order to enhance the faint nebular structures (lobes).}
  \label{fig:etacar_hst}
\end{figure*}

Eta~Car is also a massive binary star with a 5.52~yr spectroscopic periodicity (\citealt{1998A&AS..133..299D}, \citealt{2008MNRAS.384.1649D}, see also \citealt{2012ASSL..384.....D, 2016ApJ...819..131T} for a recent summary). The primary star produces a dense stellar wind with mass-loss rate $8.5 \times 10^{-4} \dot{\mathrm{M}}~\mathrm{yr}^{-1}$ and terminal velocity of 420 km $\mathrm{s}^{-1}$ \citep{2012MNRAS.423.1623G}. The secondary star has escaped direct detection so far. While not yet detected directly via its photospheric emission, the companion is thought to be responsible for time-variable ionization of the surrounding ejecta and strongly variable X-ray emission as its fast wind collides with that of the primary star wind. The secondary star is following a very highly eccentric orbit of roughly $e \simeq 0.9$ \citep{2002A&A...383..636P}. Many papers covering different wavelength regions demonstrate the variability due to the orbital periodicity of $\eta$~Car (e.g., \citealt{1997Natur.390..587C, 2001ApJ...547.1034C, 2004MNRAS.352..447W, 2004ApJ...610L.105S, 2007ApJ...660..669N, 2009A&A...493.1093F, 2011ApJ...740...80M, 2011ApJ...726..105P, 2015A&A...578A.122M}).

The sub-arcsecond region around $\eta$~Car has been resolved at visible wavelengths by \citet{1986AandA...163L...5W, 1988AandA...203L..21H, 2012ASSL..384..129W}, using  speckle interferometry and bispectrum speckle interferometry. These observations revealed three ejected knots, B, C, and D at separations between 0.1 and 0.3 arc-seconds from the central star (knot A). Eta~Car was also observed with imaging polarimetry in the H$\alpha$ line, showing a complex polarization field very close to the star \citep{1996A&A...306L..17F}. The system was also observed with the {\it Hubble Space Telescope (HST)} \citep{1995RMxAC...2...11W, 2004ApJ...605..405S}. HST/STIS spectroscopy investigated the extended wind-wind interaction zone \citep{2011ApJ...743L...3G} and discovered fossil remnants from the wind-interaction ejections in the recent past of the system \citep{2016MNRAS.462.3196G}. Proper motions of the knots $\eta$~Car B, C and D demonstrate that these knots were ejected long after the Great Eruption, most likely originating from the 1890s Lesser Eruption \citep[see][]{1995RMxAC...2...11W, 2004ApJ...605..405S, Dorland_2004, 2012ASSL..384..129W}.

The infrared images of $\eta$ Car also revealed a complex knotty structure in the inner arc\,second, 
in addition to a structure (nicknamed the "butterfly" nebula) inside the main Homunculus nebula \citep{2005A&A...435.1043C}. The "butterfly" feature coincides with a disrupted torus at the pinched waist where the polar lobes meet at the equator \citep{2005MNRAS.357.1330S, 2006ApJ...644.1151S, 2018MNRAS.474.4988S}.  In the mid-infrared, the nebula appears hollow, with a complex inner structure made of several layers \citep{2000A&A...355..155P, 2003AJ....125.1458S, 2006ApJ...644.1151S}.  These IR knots have  detected proper motions that trace back to the Great and Lesser Eruptions \citep{2011AJ....141..202A}.

Optical interferometry observations of $\eta$ Car in the near- and mid-infrared revealed the central star wind structure, appearing as a prolate wind \citep{2003A&A...410L..37V, 2005A&A...435.1043C, 2007A&A...464.1045K, Weigelt2007}.
$\eta$~Car's wind collision zone was also imaged via aperture synthesis in multiple radial velocities channels across emission lines in the K band \citet{2016A&A...594A.106W} and \citet{gravity2018}, revealing a complex and time-varying structure.
Ground-based adaptive optics imaging at visible wavelengths have recently resolved the H$\alpha$ emission region of the primary star's wind \citep{2017ApJ...841L...7W}, consistent with the range of radii expected from models \citep{2006ApJ...642.1098H}.


Imaging polarimetry of the $\eta$\,Car environment can help in determining the 3D shape of its close-by environment, as was demonstrated in \citet{1999AJ....118.1320S}. Notably, they found that the nebula around $\eta$~Car has an hourglass shape instead of a double bubble one, a shape that was subsequently confirmed in \citet{2006ApJ...644.1151S} using high-resolution long-slit spectroscopy.

In this paper, we aim to study the structure and kinematics of $\eta$ Car's ejecta by direct imaging of the innermost circumstellar material at high angular resolution.

\section{Observations and data reduction}\label{sec:obs}

We observed $\eta$~Car on February 19$^{\rm th}$, 2016 with the extreme adaptive optics Spectro-Polarimetric High-contrast Exoplanet REsearch instrument (SPHERE; \citealt{2019A&A...631A.155B}, see log of observations in Table~\ref{tab:SPHERE}),  installed on a  Nasmyth focus of the Unit Telescope 3 (Melipal) at the Very
Large Telescope (VLT) at Cerro Paranal in Chile. We made use of  the
optical imaging and polarimetric system ZIMPOL \citep[described in][]{2018A&A...619A...9S}, giving access to a theoretical 20\,mas angular resolution (for an 8\,m telescope at 656\,nm).


\begin{figure*}[htbp]
\centering
  \hspace*{-0.5cm}\includegraphics[width=0.95\textwidth,angle=0]{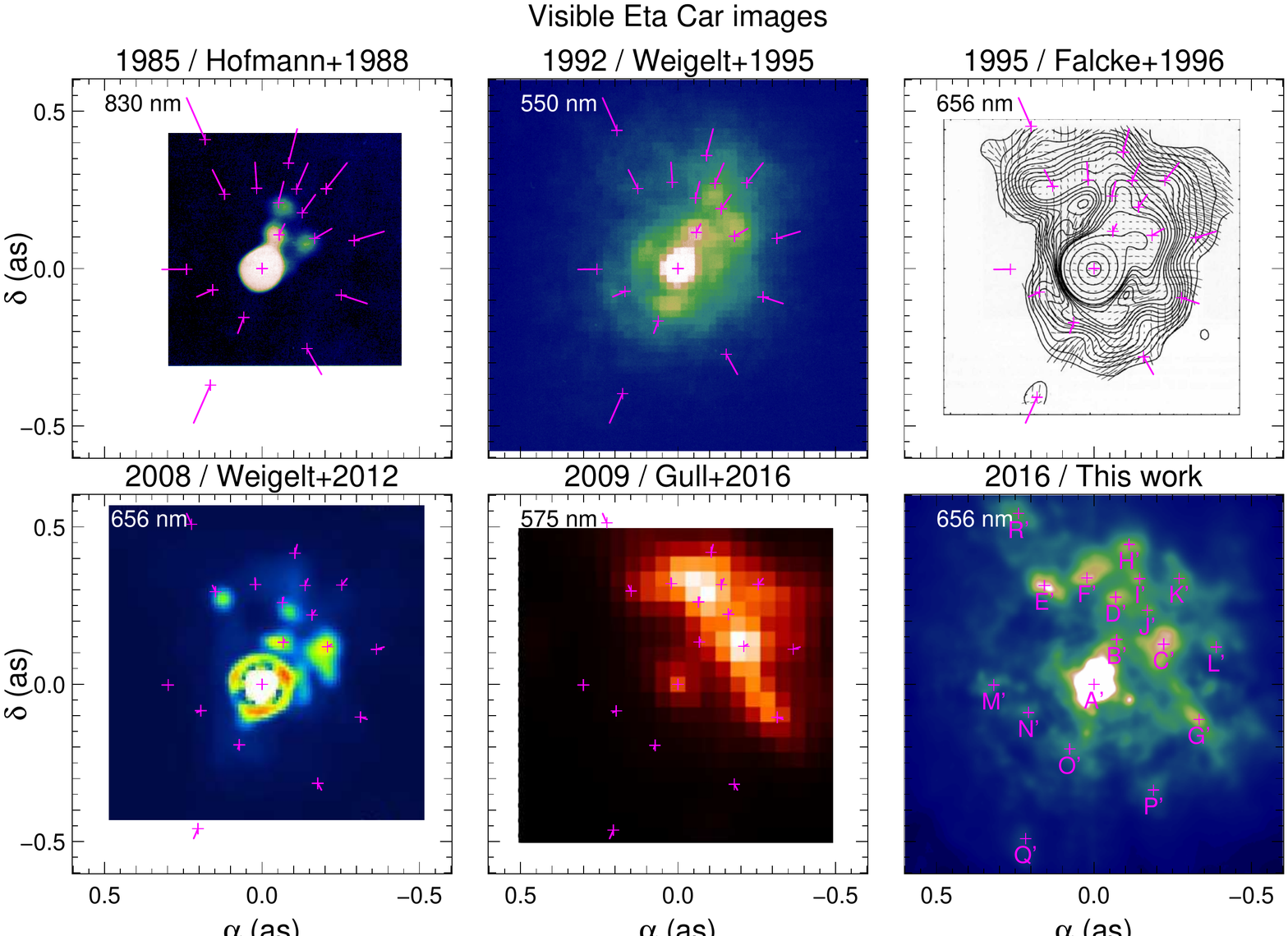}
  \vspace{0.5cm}
  \caption{{\bf From left to right and top to bottom:} Historical visible images of the inner arcsecond of $\eta$~Car. The last image (lower-right corner) shows our SPHERE/ZIMPOL deconvolved image of $\eta$~Car in the H$\alpha$ line with prominent features marked from A' to R'. The same features are displayed scaled down on top of the other images with crosses, assuming they were ejected in 1890 (see text for details). The lines represent the distance to the SPHERE position.}
  \label{fig:etacar_blobs}
 \end{figure*}

We made use of the two ZIMPOL cameras,
that enable simultaneous observations with two different filters in field-stabilized mode. We
also observed  the star HD 89682 as a measurement of the Point Spread
Function (PSF). Both $\eta$~Car and the PSF
were observed with two pairs of filters, V ($\lambda_c$=554nm, $\Delta$$\lambda_c$=80.6nm) and NR
($\lambda_c$=645.9nm, $\Delta$$\lambda_c$=56.7nm) and NH${_\alpha}$ ($\lambda_c$= 656.34nm, $\Delta$$\lambda_c$=0.97) and CntH${_\alpha}$
($\lambda_c$ =644.9nm, $\Delta$$\lambda_c$=4.1nm). 
We used the neutral
density filter ND1 to avoid saturation for this very bright object, with transmissions of 0.86, 0.83, 0.84 and 0.83$\%$ in the V, NR, H$\alpha$ and CntHa filters respectively\footnote{Link to the transmission curves: \url{https://www.eso.org/sci/facilities/paranal/instruments/sphere/inst/filters.html}}.

We reduced the data following the process described by \citet{2015A&A...578A..77K}. We first pre-processed the raw cubes using the ESO pipeline 
to produce the Stokes +Q, $-$Q, +U and $-$U frames for each camera.
We then aligned the resulting frames using custom Python routines. 
We adopted a pixel scale of 3.628\,mas.
We determined the polarimetric quantities (polarized flux, degree of linear polarization, and polarized angle) using the relation described by \citet{2015A&A...578A..77K}. 
We subsequently cropped the images to roughly $3\arcsec \times 3\arcsec$ since the useful window is not exactly centred on the target, and we focused on a clean square image for deconvolution.

We present the resulting image of $\eta$~Car and the PSF total flux (Stokes~I) in the two first columns of  Figure~\ref{fig:etacar_all} where we present the most prominent structures. We deconvolved the SPHERE total flux images using the observed calibrator as a PSF and the Lucy-Richardson algorithm, as coded in the Yorick language\footnote{Yorick is a scientific language developed by D.\ Munro \url{http://yorick.github.com}. The Lucy-Richardson algorithm in Yorick is distributed by T.\ Paumard \url{http://www.lesia.obspm.fr/perso/thibaut-paumard/yorick/}.}. Figure~\ref{fig:etacar_all} presents the  deconvolved images with 20 iterations, with a reduced field of view compared to the full image in order to focus on the knots.


\begin{figure*}[htbp]
\centering
\hspace*{-3cm}
\begin{tabular}{ccc}
\vspace*{-3cm}  \includegraphics[width=.45\textwidth]{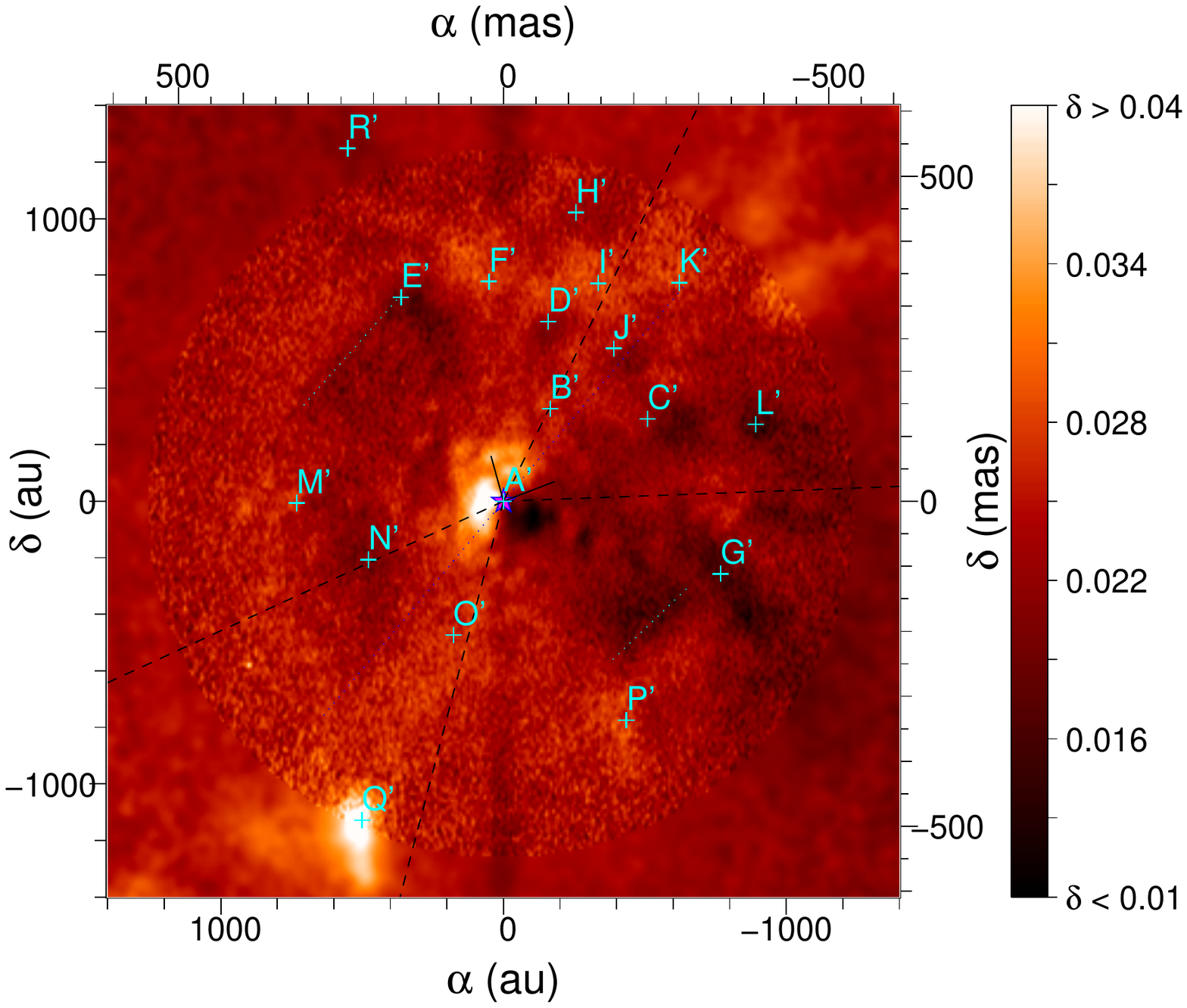} & 
\vspace*{-3cm}    \includegraphics[width=.45\textwidth]{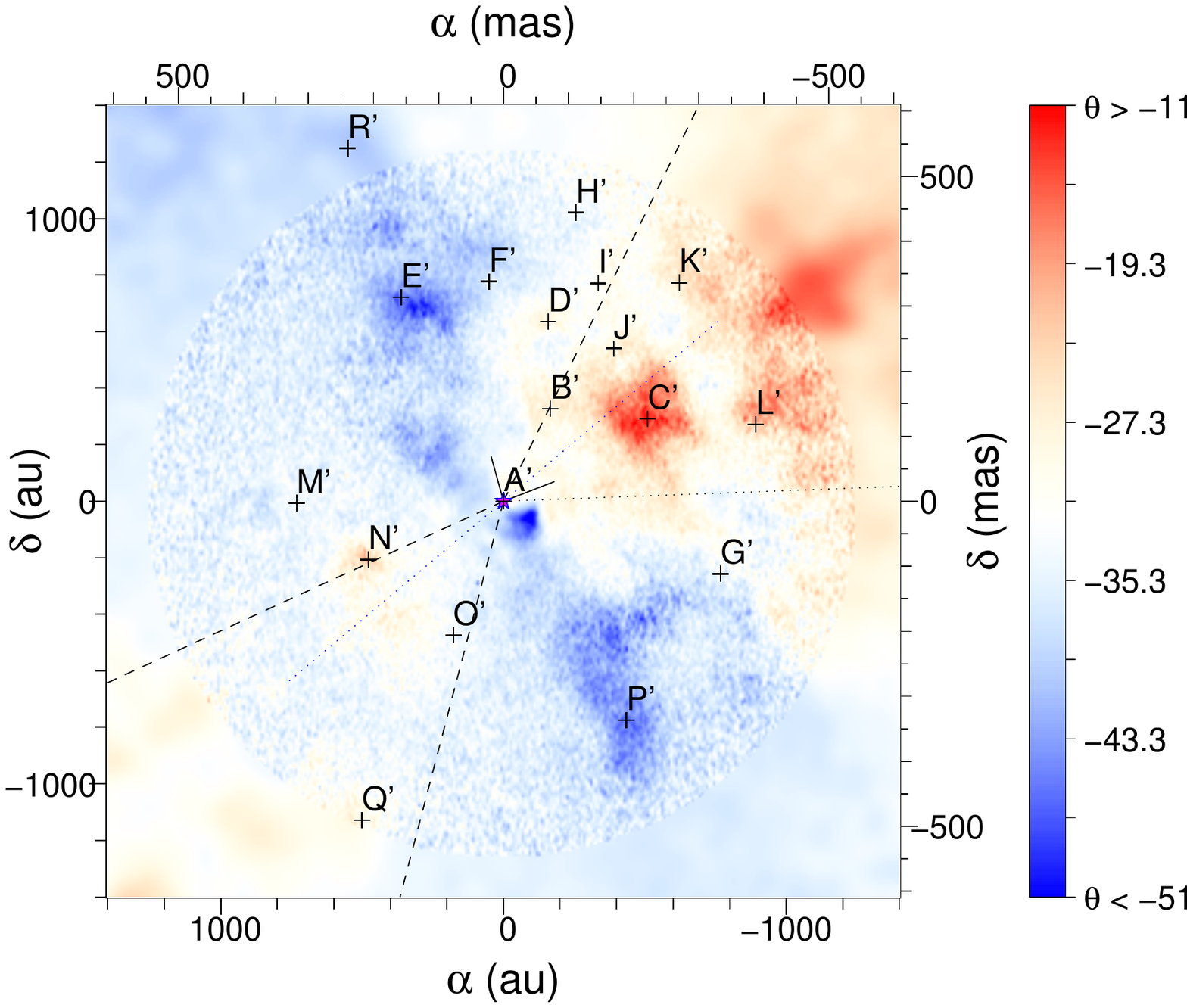} \\
\vspace*{-2.2cm}    \includegraphics[width=.45\textwidth]{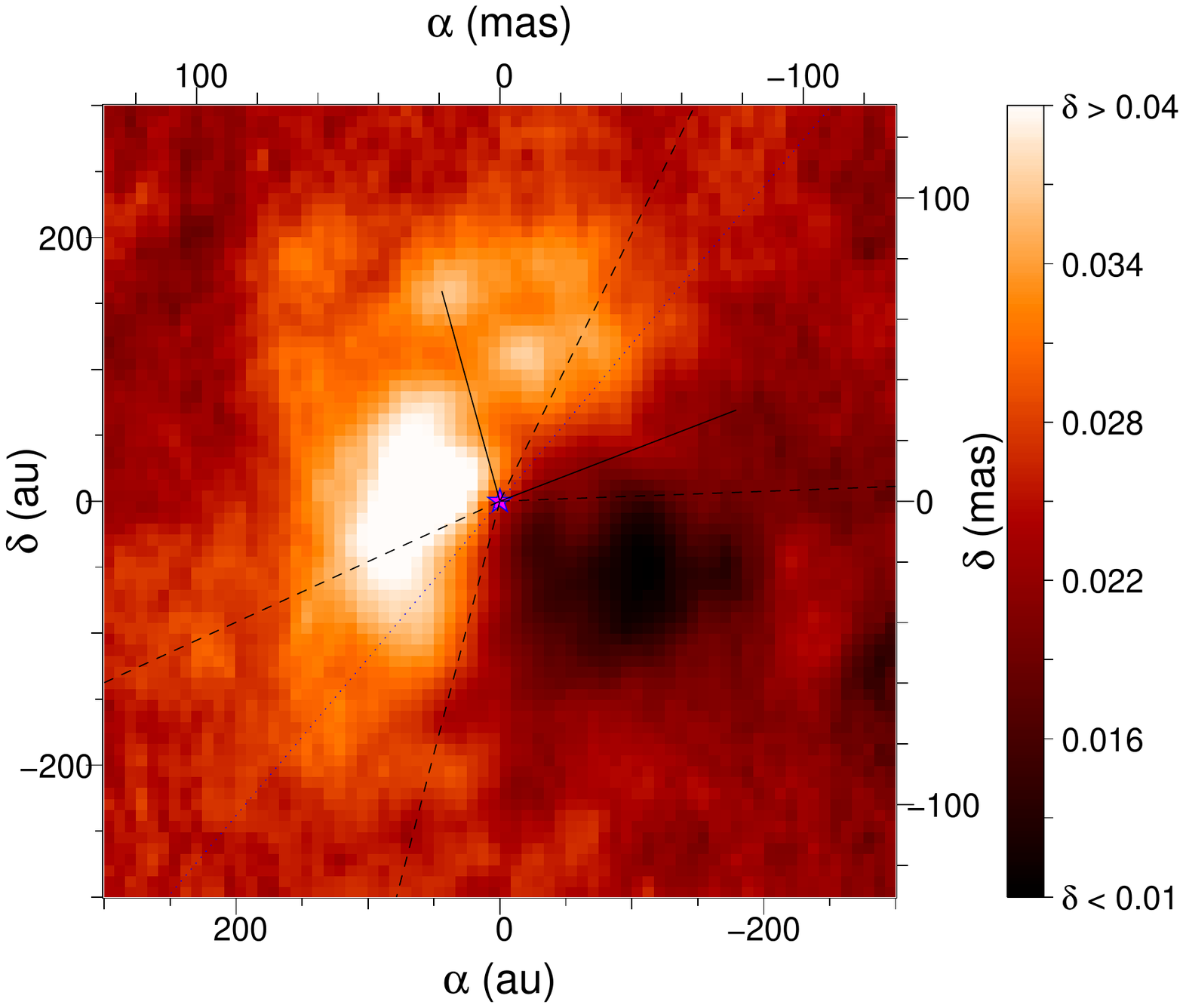} & 
\vspace*{-2.2cm}    \includegraphics[width=.45\textwidth]{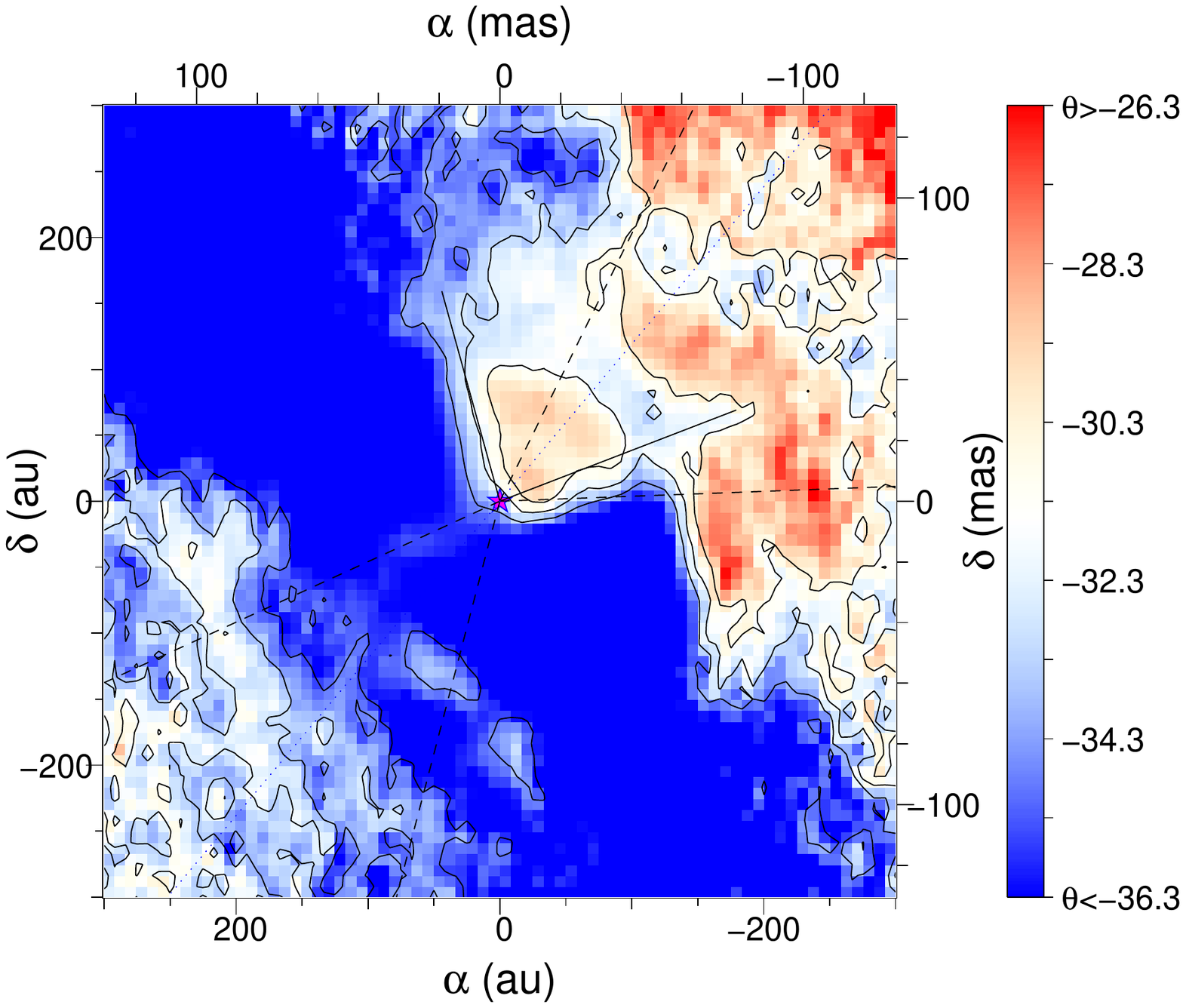} \\
  \end{tabular}
  \vspace{0.5cm}
 \caption{{\bf Top-left:} SPHERE/ZIMPOL polarisation degree map  of $\eta$~Car in the H$\alpha$ line, with the position of the 18 new knots shown in Figure~\ref{fig:etacar_hst} marked. We smoothed the image above a radius of 1\,200~au in order to keep the noise level low-enough for readability. {\bf Top-right:} Polarization angle map, with the large bi-cone (dashed lines) and the inner fan-shaped structure (solid lines) highlighted. {\bf Bottom-left:} The above map zoomed onto a $250 \times 250$~mas ($300 \times 300$~au) region. {\bf Bottom-right:} The polarization angle map in a $250 \times 250$~mas region around the central star, plotted between $-$25 and $-$35$^\circ$. Contours are at $-$31, $-$33 and $-$35$^\circ$.}
  \label{fig:etacar_polar}
 \end{figure*}

We also performed a Lucy-Richardson deconvolution with 80 iterations, a PSF subtraction, and a Wiener deconvolution, shown in Figure \ref{fig:etacar_all_end}, to check similarities and differences that depend on the restoration technique and parameters.
Deconvolution artefacts range from  bright or dark "rays", horizontal "stripes", or "ghost" of the PSF. They correspond to mismatched telescope spider arms between the star and the PSF (we used SPHERE in "field-stabilized" mode and not "pupil-stabilized" mode), uncorrected detector pattern, and finally AO behaviour difference between the star and calibrator, respectively. We indicate by arrows the strongest deconvolution artefacts described above in Figure~\ref{fig:etacar_hst}.

In Figure~\ref{fig:etacar_hst}, we show only the H$\alpha$ filter image for the useful SPHERE field of view, compared to an archival H$\alpha$ {\it HST/ACS} image from 2003. HST (a 2.4\,m telescope) has a typical angular resolution of 70\,mas at 656\,nm. 
Figure~\ref{fig:etacar_blobs} shows a comparison of our H$\alpha$ image with previous detections of the $\eta$~Car knots. We will discuss this comparison in the next section.

In Figure~\ref{fig:etacar_polar}, we present the polarization maps (polarization degree and polarization angle) in the H$\alpha$ line. The polarization map presents a marked East-West ``two lobes'' artefact in the polarization degree in the 100\,mas area close to the central star. This artefact is a known problem of the method used to compute the polarized quantities with a singularity at the centre, but this issue does not affect the present study.




\section{Results}\label{sec:results}

\subsection{More knots in the core of $\eta$~Car}\label{sec:blobs}

 The total flux (Stokes I) images (Figure \ref{fig:etacar_all} and Figure~\ref{fig:etacar_hst}) show a wealth of information: one can recognize prominent features of the inner arcseconds of the Homunculus nebula, especially the inner part of the South-East lobe, seen as a diagonal arc spanning the image (marked with a dashed line in Figure~\ref{fig:etacar_hst}). 
 
We also identified 18 separate bright features or knots, that can be seen in  the deconvolved images in all four filters (Fig.~\ref{fig:etacar_all}). We label these knot alphabetically from A' to R'.

Figure~\ref{fig:etacar_blobs} shows our image marked with the detected features in the lower-right panel. The same figure presents historical images of $\eta$~Car, where we scaled the SPHERE features to the previous epochs, adopting an 1890 ejection date, i.e. a scale factor of 0.82 between 1985 and 2016 (as illustrated by the lines in the figure~\ref{fig:etacar_blobs} panels). Changing this date by 10s of years (e.g. adopting 1847 as ejection date) does not change that scale factor significantly. This knots ejection date is currently reported between 1880 and 1890 \citep{2012ASSL..384..129W, 2017MNRAS.471.4465S}, although the actual ejection date of the Homunculus is thought to be between 1843 and 1847 \citep{2001ApJ...548L.207M, 2011AJ....141..202A, 2017MNRAS.471.4465S}.



\begin{figure}[htbp]
  \centering
\vspace*{-1cm}
 \hspace*{-1.cm}\includegraphics[width=1\columnwidth]{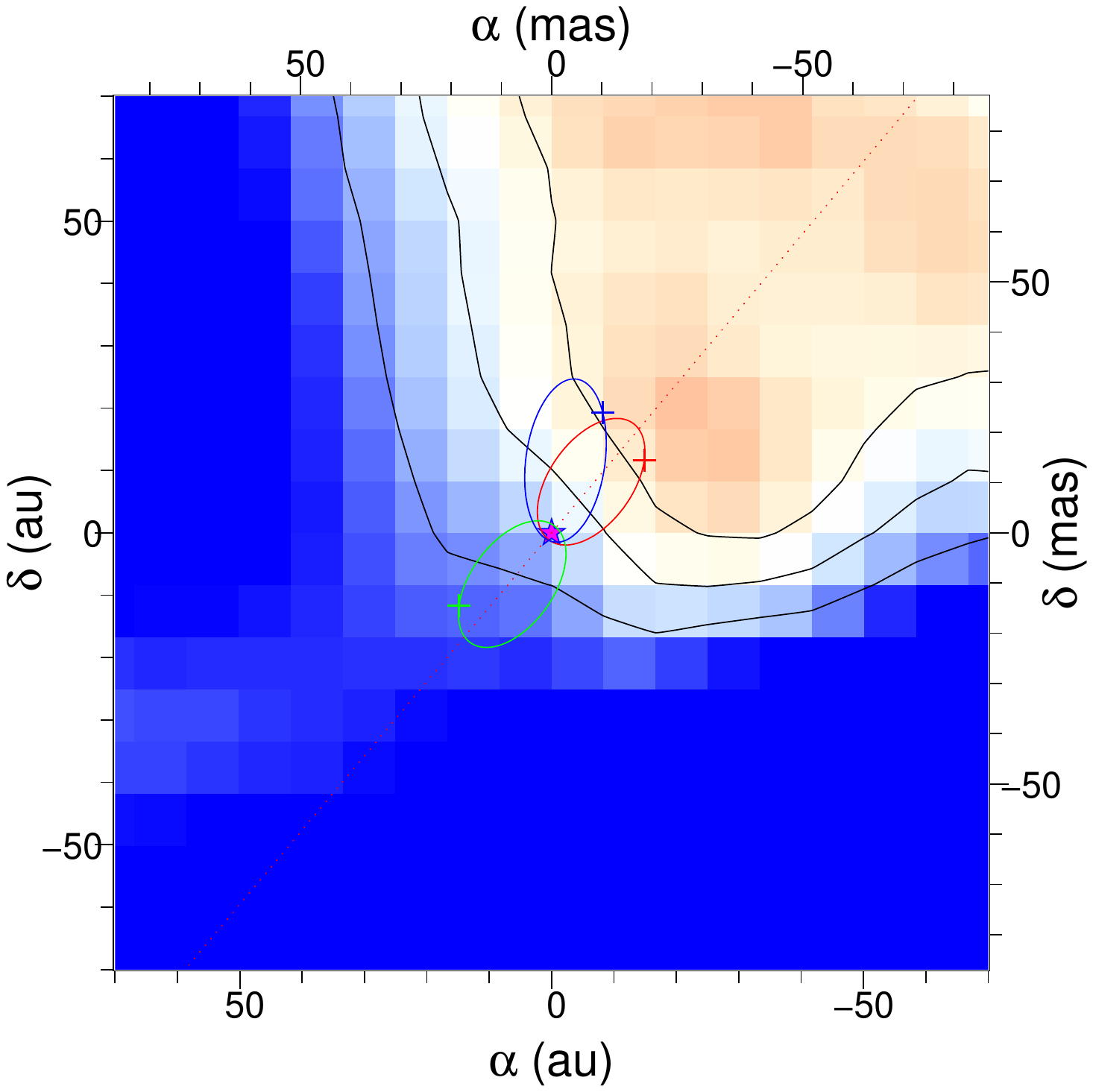}
\vspace*{-4cm}
\caption{ Angle of polarization map with same colour scale as in Figure~\ref{fig:etacar_polar}. The orbit from \citet{2012MNRAS.420.2064M} (red), \citet{2016ApJ...819..131T} (blue) and \citet{2018ApJ...858..117K} (green) are shown, together with the predicted position of the companion star at the epoch of observation. }
  \label{fig:etacar_orbit}
\end{figure}

Figure~\ref{fig:etacar_blobs} shows that knots B' and C' likely correspond to structures seen in the same filter in 1995 and 2008. They may correspond to a scaled-up version of the knots B and C of \citet{1988AandA...203L..21H}, seen also in 1992. One must be cautious, though, about their precise positions at different wavelengths and in different filters, because at visual wavelengths these knots are seen as emission-line structures (i.e. an intrinsic emission, best seen in narrow-band filters, and also reflected emission, best seen in wideband filters). Their size and centroid position depends on which filter is used.

Knot D' does not match well the knot D in 1995 and 2008 \citep[and][]{1988AandA...203L..21H}, but appears to be fragmented into three knots (D', I', and J'). It may not seem impossible that the structure of the knots evolved from 2008 to 2016. However, it is unclear to us what caused the time evolution of the morphology, and velocity information would be useful for interpreting this, an information that we do not have with the SPHERE images.


The two bright knots E' and F' are undocumented, but were already present in historical images, clearly seen in \citet{1996A&A...306L..17F} and \citet{2004ApJ...605..405S}, and maybe also in \citet{2016MNRAS.462.3196G}, and knot E' may be seen in \citet{2012ASSL..384..129W}. We note here that the knot E' is seen the in H$\alpha$ filter images (1995, 2008, 2016), and not in the other filters (1988, 1992). E' may be seen in the 2009 images but that is an unclear detection.

The knot marked as G' was already identified by \citet{2016MNRAS.462.3196G} and indeed looks like an elongated feature or a bar. It probably corresponds to the elongation seen in knot C' of \citet{2012ASSL..384..129W}.

These condensations near the star are highly variable in their emission line fluxes, changing with the binary orbital cycle and also showing long-term changes in intensity, with some emission lines fading as others brightening \citep{2000AJ....120..920S}. 
The emission line knots showing the strongest flux variability in the orbital cycle appear to be in a flattened, perhaps equatorial distribution with a radial gradient in ionization level \citep{2000AJ....120..920S}.
 
We note that knot O' might be a fainter symmetric counterpart of knot B' relative to A'. The same applies to the pairs N' / C', and P' / E'. The other knots do not appear to have a symmetric counterpart.

\subsection{Polarization degree and angle}\label{sec:polar}

Figure~\ref{fig:etacar_polar} shows the polarization degree and polarization angle maps of $\eta$~Car in the inner arcsecond. 
The polarization degree map exhibits an overall low degree of polarization, between 0\% and 5\%. This is far less than reported in \citet{1996A&A...306L..17F} and could be related to the dissipating dust environment \citep{2019MNRAS.484.1325D}.
Another hypothesis may be the evolution of the knots: for instance, the diameter and the flux of each knot may have changed compared to the 1985 and 2008 images.

Knots B', C', D', E', G', J', L', N', and R' mark low polarization spots in the polarization degree map (likely due to forward scattering, i.e. in front of the star), while knots F', H', I', K', M', O', P', Q' mark high polarization spots in the same map (due to 90 degrees scattering).


The polarization degree falls close to zero in a line oriented by $\sim 45^\circ$, i.e. roughly in the equatorial region of the Homunculus nebula. In the same region, the polarization angle shows a roughly constant $-45^\circ$ value.
We note that knots E' and P' lie in this region. Knot E' is in a low polarization state and knot P' in a higher polarization state, and both share the same polarization angle.


The polarization degree and polarization angle show a structured bi-conical shape extending from the inner hundred au from the star up to several thousand au. One can best see this bi-cone structure in the polarization angle map, where the angles range from $-$30 to $-$10$^\circ$ (Figure~\ref{fig:etacar_polar}).
\citet{1996A&A...306L..17F} already detected this bi-cone, albeit with fewer details than with SPHERE.


In the inner 300~au region from the star (two lower panels of Figure~\ref{fig:etacar_polar}), we detect a one-sided fan-shaped structure, offset from the  bi-cone structure mentioned above. This fan-shaped structure spans from 20~au up to 200~au, with an opening angle of 86$^\circ$ and an orientation of $-$26$^\circ$. It is best seen in the H$\alpha$ line maps and we highlight it in Figure~\ref{fig:etacar_polar} lower panel polarization angle map with contours at $-$31, $-$33 and $-$35$^\circ$.

\section{Discussion}

\subsection{A previously seen inner bi-cone structure}

The bi-cone structure appearing as red structures in the upper-right panel of  Fig.~\ref{fig:etacar_polar} and labelled as dashed lines shows a more substantial polarization degree than the other regions of our maps. Such a high polarization degree may result from scattering by dust in the inner regions of the Homunculus nebula.

The South-Eastern cone axis is oriented at position angle (PA) $320^\circ$ while the North-Western cone axis is at PA $303^\circ$. They are roughly aligned with the Homunculus main axis (PA $310^\circ$), though with an approximate offset of $10^\circ$.

Opening angles are the following: SE cone: $51^\circ$, NW cone: $62^\circ$, i.e. similar, but again with a difference of $10^\circ$.



We also can see an equatorial flattened structure of low polarization degree, that presents two "bars" seen in polarization degree and polarization angle and marked in the upper-left panel of figure~\ref{fig:etacar_motion}.

\begin{figure*}[htbp]
  \centering
 \includegraphics[width=18cm]{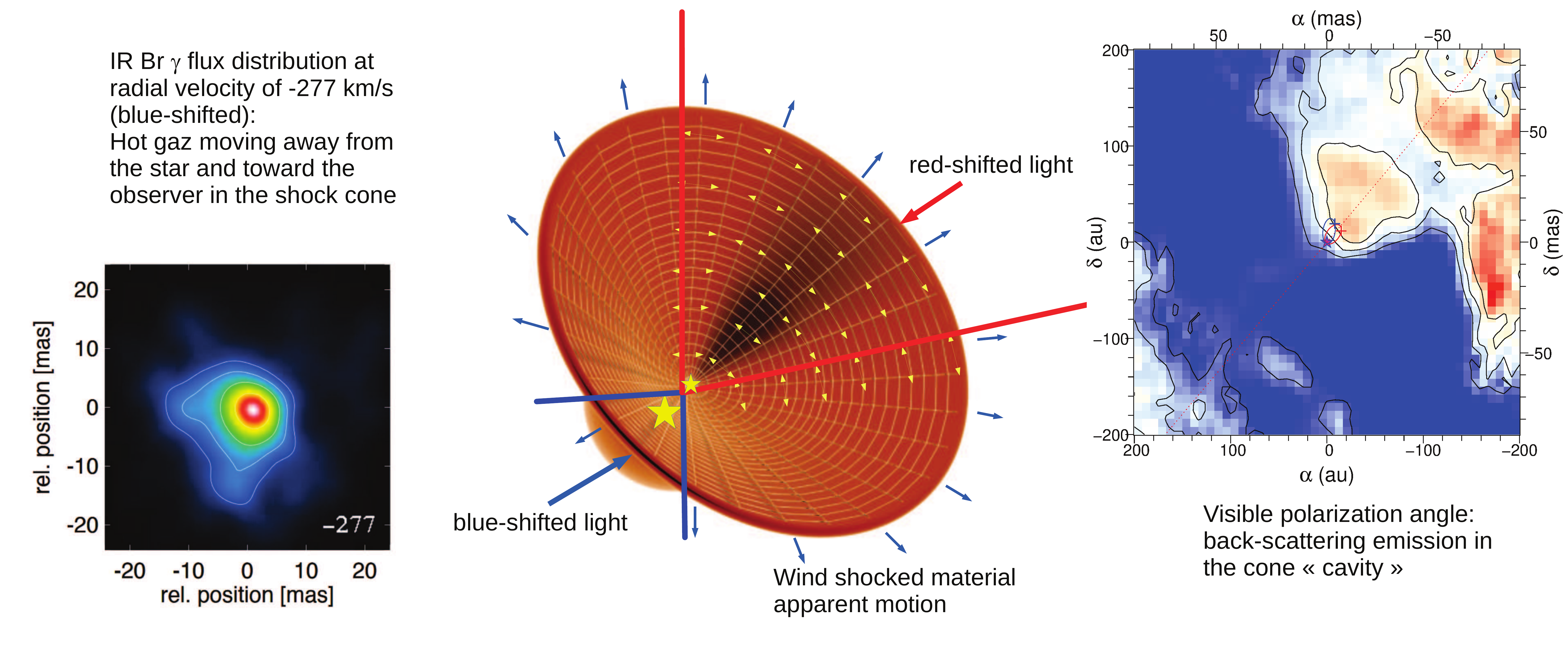}
  \caption{Sketch showing the expected features observed close to the star. The left panel is a reproduction of Figure 6 from \citet{2016A&A...594A.106W}. The middle panel is from the same figure with added annotations, showing the blue-shifted emission coming from the SE part of the cone (between the two blue lines), and the red-shifted emission, that has an inclined back-scattered polarization (yellow arrows). The right panel is the fan-shaped feature seen in the SPHERE polarization angle map.}
  \label{fig:etacar_sketch}
\end{figure*}

\subsection{Expanding Knots that are more numerous than seen before}

Knot D' could have been fragmented since the last recorded epoch, or we take profit of the unprecedented angular resolution provided by SPHERE ($\approx20$\,mas) to resolve knot D into three knots.

This means that over the last 21 years, the knots B', C', D', E' and F' have likely varied in relative intensity, knot B' being fainter after 2008--2009, knot E' being absent before 1995 and  getting brighter with time. The fact that we detect apparently fragmented knots means that these knots may not form self-sustained structures, but rather overdensities in a turbulent medium.

Post-outburst turbulent layers growing in time have been invoked by several authors \citep{10.1007/BFb0106406} to explain the cauliflower appearence of $\eta$~Car's homunculus. Indeed, the mottled appearance of the Homunculus polar lobes has been attributed to Vishniac instabilities \citep{2013MNRAS.429.2366S}.




In the many new knots detected here, some appear to be the symmetrical counterpart of other knots, for the first time. This means that the ejection of knots may have been centrosymmetric, or axisymmetric, and that the newly detected knots in this work were too faint to be previously detected.

Only 3 knots have a significant number of positions measured: B', C' and D' (including the \citealt{2004ApJ...605..405S} points for C' and D'). A simple linear regression gives an ejection date of 1884 for knot B', 1918 for knot C'  and 1900 for D', as explained in appendix~\ref{sec:appendix}. One can note here that epochs of ejection from proper motions do not depend on the projection angle on the sky. The consequence is that the inferred date will not change with inclination angle. In any case, those dates do depend on linear motion. These 3 dates are dependent on which filter is used, and especially we get significantly different dates when using only the H$\alpha$ images (3 epochs). 



\subsection{Wind collision zone}

The orbit orientation of $\eta$~Car's companion star has been the subject of many discussions. Most recent works on the determination of the orbit are \citet{2012ApJ...746L..18M, 2012MNRAS.420.2064M, 2014MNRAS.442.3316S, 2016ApJ...819..131T, 2018ApJ...858..117K}. In Figure~\ref{fig:etacar_orbit} we compare the polarization angle map obtained with SPHERE with three potential orbits of the companion to the same scale, taken from \citet[][red]{2012MNRAS.420.2064M}, from \citet[][blue]{2016ApJ...819..131T}, and from \citet[][green]{2018ApJ...858..117K}. We mark with a "+" the expected position of the companion star at the date of observation. In two cases, the position of the companion star roughly corresponds to the tip of the fan feature presented previously. For the orbit proposed by \citet{2018ApJ...858..117K}, the position of the companion does not match that fan tip. Therefore, this fan-like feature may be a part of the cone of the wind collision zone between the two stars. The companion star orbit apoastron must therefore be located close to the tip of that fan-shaped structure. We expect to see changes in this feature when the system gets closer to the periastron passage.

\citet{2016A&A...594A.106W} detected a fan-like structure opening to the SE in the blue-shifted region of the Br$\gamma$ line (-277\,km/s).  We present in Fig.~\ref{fig:etacar_sketch} the sketch of \citet{2016A&A...594A.106W} annotated in an effort to explain what we see in the polarization angle map of SPHERE: while the near-IR blue-shifted Br$\gamma$ images probe column density of gas, the polarization degree and angle indicate back-scattering of the star's light on the surface of the denser shock region, away from us (and likely red-shifted). In the blue-shifted region of the Br$\gamma$ line, the SE region of the sketched conical cavity of Figure~\ref{fig:etacar_sketch} is brighter, because the gas moves from the centre of the cavity to the edge of the cavity, and towards the observer.




The opening angle of the fan-shaped structure is $86^\circ$. This is smaller than the estimates of the WCZ opening angle from \citet{2012MNRAS.423.1623G} of 100-140$^\circ$. However, in the sketch shown in Fig~\ref{fig:etacar_sketch}, one can see that because of projection effects, the fan-shaped structure seen in the polarimetric signal can have a smaller opening angle than the WCZ cone opening angle. Future polarimetric models are needed to further investigate this finding, with our observations being potentially able to offer important constrains on the properties of the wind-collision zone.

\section{Conclusions}

We have shown SPHERE/VLT images of $\eta$~Car in visible continuum wavelengths and the H$\alpha$ emission line both in polarized light.

These new images exhibit a better angular resolution than previous images from HST \& VLT(NACO), with $\approx20$\,mas details clearly seen in the images. Our images show more knots than before, including those ones seen as faint symmetrical counterparts of previously known ones.

Our polarization maps display similar bi-cone features as was already seen in \citet{1996A&A...306L..17F}.
In addition, a one-sided NW fan-shape structure with an opening angle of $86^\circ$ can be detected in the polarization degree and polarisation angle SPHERE maps, very close to the central star. This new structure can be naturally matched with some orbital hypotheses (with $\omega$ close to $270^\circ$) for the central binary system, placing the companion star at apoastron at the tip of that structure. Our proposition is that this fan-shape structure corresponds to back-scattering of the star's light directly on the denser WCZ of the system.

We expect to see changes in this feature when the system gets closer to the periastron passage. 
Therefore, we call for more observations with the exquisite angular resolution of SPHERE in the future. These observations would be the most effective accompanied by simultaneous interferometric imaging.

\begin{acknowledgements}
This work has made use of the SPHERE Data Centre, jointly operated by
OSUG/IPAG (Grenoble), PYTHEAS/LAM/CeSAM (Marseille), OCA/Lagrange
(Nice) and Observatoire de Paris/LESIA (Paris) and supported by a
grant from Labex OSUG@2020 (Investissements d’avenir – ANR10 LABX56).
This project has received funding from the European Union’s Horizon 2020 research and innovation programme under the Marie Sk\l{}odowska-Curie Grant agreement No. 665501 with the research Foundation Flanders (FWO) ([PEGASUS]$^2$ Marie Curie fellowship 12U2717N awarded to MM). This work benefitted from support of ASHRA and PNPS, both from the CNRS. We are greatful to A. Lantieri for his wise advises on image deconvolution techniques.
\end{acknowledgements}

%
%

\bibliographystyle{aa} 
\bibliography{biblio} 

\appendix
\label{sec:appendix}

\section{Position and angle of $\eta$~Car's knots}

\begin{table*}[htbp]
\caption{$\eta$~Car knots at different epochs. We used the literature knot position when available, and recomputed them based on the best quality printout available otherwise (the digital versions of several of the published images are lost. These new computations are marked with bold face).}
\label{param}
\centering
\begin{tabular}{c|c|c|c|c|c|c|c|c|c|c|c|c|}
$\lambda$ & A- & B & C & D & E & F & G & H & I  \\
\hline
830\,nm & 1985.047 & 0.114" & 0.177" & 0.211" &-&-&-&-&-\\
 Speckle$^1$   &  & -20.0$^\circ$ & -63.5$^\circ$ & -24.5$^\circ$ &&&&&\\
\hline
550\,nm & 1992.925 & 0.123"      & 0.208"      & 0.241"      &-& {\bf0.318"} &-&-&-  \\
  HST$^2$   &      & -26$^\circ$ & -56$^\circ$ & -24$^\circ$ & & {\bf -4$^\circ$} &&&  \\
\hline
H$\alpha$ (656\,nm) & 1995.283 & {\bf 0.163"} & {\bf 0.169"} & {\bf 0.245"} & {\bf 0.295"} & {\bf 0.297"} & {\bf 0.254"} & {\bf 0.354"} &-\\
Speckle$^3$ && {\bf-26$^\circ$} & {\bf-55$^\circ$} & {\bf-22$^\circ$} & {\bf31$^\circ$} & {\bf9$^\circ$} & {\bf-112$^\circ$} & {\bf-10$^\circ$} &\\
\hline
656\,nm & 2008.03 & {\bf0.146} & {\bf0.235} & {\bf0.251} & {\bf0.300} &-&-&-&- \\
Speckle$^4$ && {\bf-19$^\circ$} & {\bf-60$^\circ$} & {\bf-21$^\circ$} & {\bf 24.2$^\circ$} &&&& \\
\hline
575.619\,nm & 2009 & - & {\bf0.237"} & {\bf0.315"} & -& -& {\bf0.299"} &-&-\\
HST$^5$ && & {\bf-52$^\circ$} & {\bf-15$^\circ$} &&& {\bf-105$^\circ$} && \\
\hline
656\,nm & 2016.219 & {\bf0.159"} & {\bf0.255"} & {\bf0.285"} & {\bf0.351"} & {\bf0.338"} & {\bf0.352"} & {\bf0.458"} & {\bf0.365"} \\
SPHERE$^6$&& -27$^\circ$ & {\bf-60$^\circ$} & {\bf-14$^\circ$} & {\bf27$^\circ$} & {\bf4$^\circ$} & {\bf-109$^\circ$} & {\bf-14$^\circ$} & {\bf -24$^\circ$}\\
\hline
\end{tabular}
\vspace{0.5cm}\\
\begin{tabular}{c|c|c|c|c|c|c|c|c|c|c|}
$\lambda$ &  A- & J & K & L & M & N & O & P & Q & R \\
\hline
830\,nm & 1985 &-&-&-&-&-&-&-&-&-\\
Speckle$^1$ &&&&&&&&&&\\
\hline
550\,nm & 1992 &-& {\bf0.388"}&-&-& {\bf0.207"} & {\bf0.124"} &-&-&- \\
HST$^2$    && & {\bf-40$^\circ$} & & & {\bf114$^\circ$}  & {\bf163$^\circ$}  &&& \\
\hline
H$\alpha$ (656\,nm) & 1995 &-&-&-&-& {\bf0.178"} & {\bf0.170"} &-& {\bf0.450"} &-\\
Speckle$^3$ & & & && & {\bf125$^\circ$} & {\bf150$^\circ$} && {\bf156$^\circ$} &\\
\hline
656\,nm & 2008 &-&-&-&-&-&-&-&-&- \\
Speckle$^4$ &&&&&&&&&&\\
\hline
575.619\,nm & 2009 &-&-&-&-&-&-&-&-&- \\
HST$^5$ &&&&&&&&&&\\
\hline
656\,nm & 2016 & {\bf0.290"} & {\bf0.431"} & {\bf0.406"} & {\bf0.318"} & {\bf0.226"} & {\bf0.220"} & {\bf0.386"} & {\bf0.537"} & {\bf0.593"} \\
SPHERE$^6$&& {\bf-36$^\circ$} & {\bf-39$^\circ$} & {\bf-73$^\circ$} & {\bf-91$^\circ$} & {\bf113$^\circ$} & {\bf160$^\circ$} & {\bf-151$^\circ$} & {\bf156.1$^\circ$} & {\bf23.8$^\circ$} \\
\hline
\multicolumn{4}{l}{\footnotesize References:}\\
\multicolumn{3}{l}{\footnotesize 1: \citet{1988AandA...203L..21H}} & \multicolumn{3}{l}{\footnotesize 4: \citet{2012ASSL..384..129W}}\\
\multicolumn{3}{l}{\footnotesize 2: \citet{1995RMxAC...2...11W}} & \multicolumn{4}{l}{\footnotesize 5: \citet{2016MNRAS.462.3196G}}\\
\multicolumn{3}{l}{\footnotesize 3: \citet{1996A&A...306L..17F}} & \multicolumn{4}{l}{\footnotesize 6: This work}\\
\end{tabular}
  \label{tab:etacar_blobs_ast}
 \end{table*}
 
We assess here the position and motion of the $\eta$~Car knots, considering the good images we could achieve with SPHERE. As stated in the main text of the paper, some of the new knots discovered by SPHERE are visible -- though not mentioned -- in previous works. We therefore give here a rough assessment of their respective position to the central star in Table~\ref{tab:etacar_blobs_ast}. For all these works, we printed the best available quality images (some of them scanned from an old rendering), and made two astronomers point at the visible structures with a pen. The position of each knot was retrieved using a ruler and converted into separation and angle, and the average of both position was computed.

\begin{figure}[htbp]
  \centering
  \vspace{-2cm}
 \includegraphics[width=1\columnwidth]{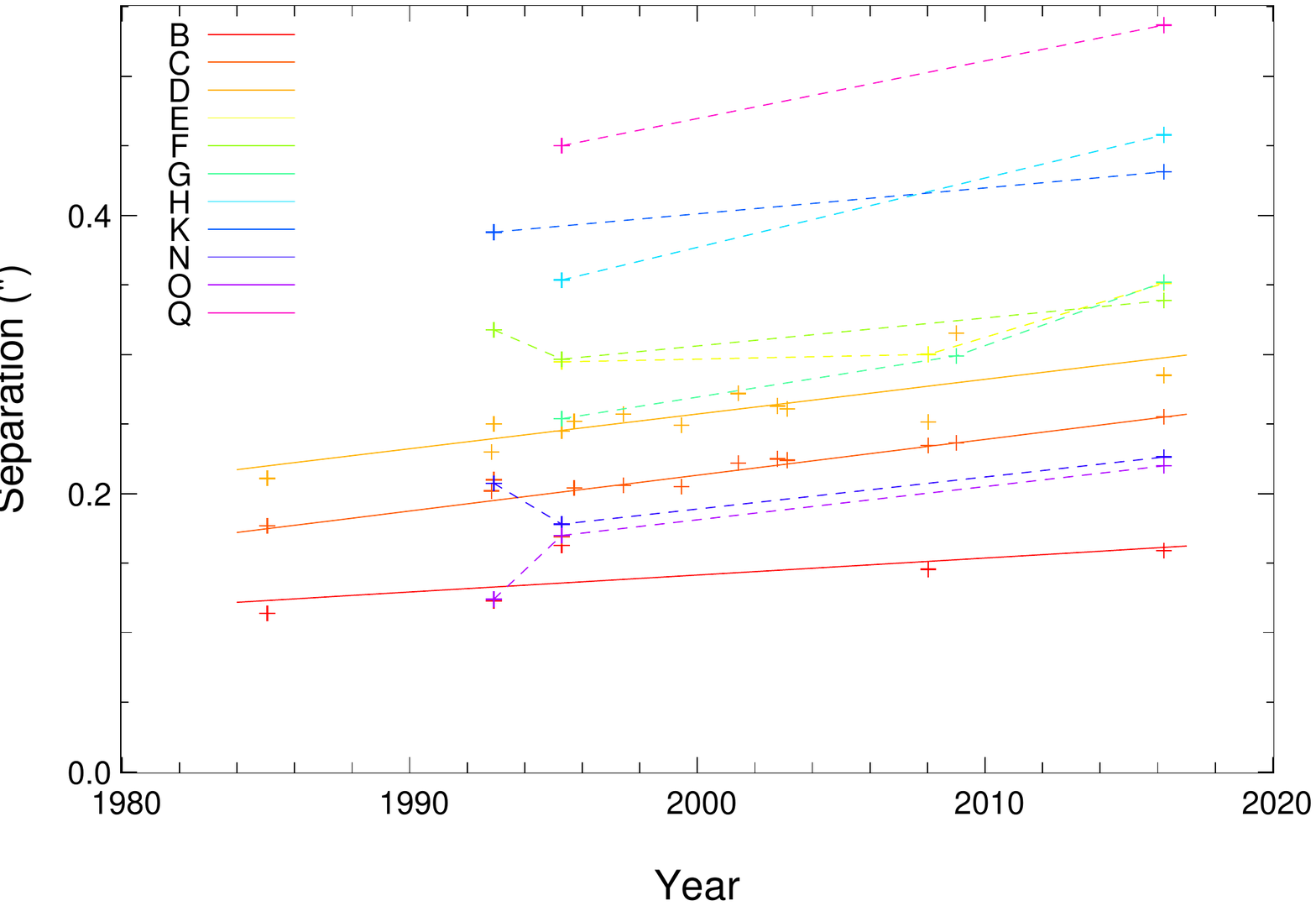}
   \vspace{-4.5cm}
  \caption{$\eta$~Car knots separation as a function of year of observation. The knots for which there is less than 3 observations are represented as simple linked lines. The knots C and D have additional measurements from \citet{2004ApJ...605..405S}. A simple linear regression is shown as a dashed line for knots B, C and D, that have more than 3 observations.}
  \label{fig:etacar_motion}
\end{figure}

The separation as a function of time for each knot is shown in Figure~\ref{fig:etacar_motion}, including the very precise measurements from \citet{2004ApJ...605..405S} for knots C and D.

One can see that all knot appear to flee from the central star, as was already pointed out previously, at an average rate of 2.8\,mas/year, but with a large scatter of 1.3\,mas/year. We can use a naive linear regression to compute an ejection date  for knot B': 1884, C': 1918, and D': 1900. The whole set of knots for which we can measure position at least for 2 points in time give ejection dates which are scattered between 1800 and 1950. This is mainly due to the fact that we compare objects with different filters. If we restrict to H$\alpha$ filter observations (i.e. 1995, 2008 and 2016), we only have 3 epochs left, and the ejection dates become C': 1954, and D': 1861, while B' does not seem to move in the 3 H$\alpha$ images.
This results from the large uncertainty on the epochs of ejection from astrometry data obtained close to the resolution limit of the telescopes. A new SPHERE image with a time separation of a few years to the ones presented here, and obtained with pupil-tracking would help positioning with better precision each knot, and infer a proper motion for all the 18 knots presented here.
We also find an expansion rate independent of position angle.

\section{Image restoration using other techniques}

\begin{figure*}[htbp]
\centering
  \hspace*{-1cm}\includegraphics[height=19cm, angle=0]{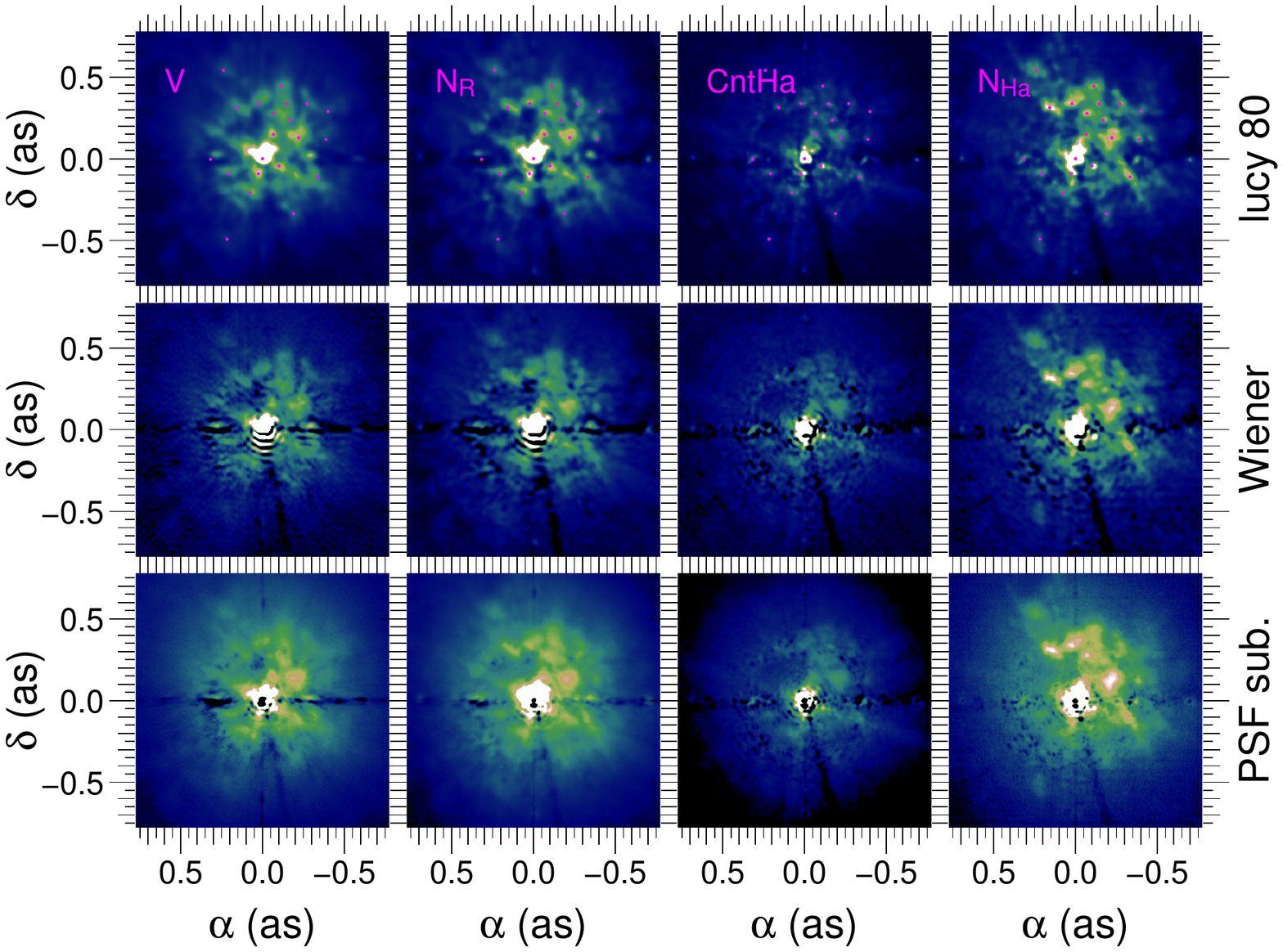}
  \vspace*{-4cm}\caption{{\bf From left to right:} SPHERE images of $\eta$~Car in the V band ($\lambda=554$\,nm), R band ($\lambda=645.9$\,nm), H$\alpha$ continuum filter ($\lambda=644.9$\,nm), H$\alpha$ filter ($\lambda=656.34$\,nm). {\bf From top to bottom:} lucy-deconvolved images of $\eta$~Car with 80 iterations, Wiener-deconvolved images, and finally PSF subtraction. We mark the identified features with red dots on the Lucy 20 line.
}
  \label{fig:etacar_all_end}
\end{figure*}

We present in Figure \ref{fig:etacar_all_end} the other techniques used to enhance and identify the  knots in the $\eta$ Car inner arcsecond environment. The figure presents a Lucy-Richardson deconvolution using more iteration than the main presented image (80 iteration instead of 20), a simple Wiener deconvolution, and a PSF subtraction. One can see that the detected knots are all present in the images .

\end{document}